\documentclass[aps, prb, reprint, superscriptaddress]{revtex4-2}

\usepackage[utf8]{inputenc}
\usepackage[T1]{fontenc}
\usepackage{newtxtext}
\usepackage{newtxmath}
\usepackage[scaled=0.92]{helvet}
\usepackage[english]{babel}
\usepackage{microtype}
\usepackage{xspace}
\usepackage{amsmath, amsfonts}
\usepackage{bm}
\renewcommand*{\epsilon}{\varepsilon}
\renewcommand*{\theta}{\vartheta}
\renewcommand*{\rho}{\varrho}
\renewcommand*{\phi}{\varphi}

\renewcommand{\vec}[1]{\bm{#1}}
\newcommand*{\hamil}{\mathcal{H}}

\newcommand*{\tensorial}[1]{\mathcal{#1}}
\newcommand*{\op}[1]{\hat{#1}} \newcommand*{\vecop}[1]{\hat{\vec{#1}}} \newcommand*{\iu}{\mathrm{i}}  \usepackage[version=4]{mhchem}
\usepackage{leftindex}
\usepackage{braket}
\usepackage[range-units=single, list-units=single, detect-all]{siunitx}
\DeclareSIUnit\rydberg{Ry}
\usepackage{graphicx}
\usepackage[autostyle]{csquotes}
\usepackage{booktabs}
\usepackage[breaklinks, hidelinks]{hyperref}

\providecommand*{\fig}{Fig.\@\xspace}

\graphicspath{{figures/}} 
\newcommand*{\acronym}[1]{\textsc{\MakeLowercase{#1}}} 

\newif\ifusesections 
\usesectionsfalse

\newcommand*{\affilUKN}{  Fachbereich Physik, 
  Universität Konstanz, 
  D-78457 Konstanz, 
  Germany} 
\newcommand*{\affilWRC}{  Department of Theoretical Solid State Physics,  
  Institute for Solid State Physics and Optics, 
  HUN-REN Wigner Research Centre for Physics, 
  H-1121 Budapest, 
  Hungary} 
\newcommand*{\affilBME}{  Department of Theoretical Physics, Institute of Physics, 
  Budapest University of Technology and Economics, 
  Műegyetem rkp.\ 3., 
  H-1111 Budapest, 
  Hungary}

\begin{document}

\title{Quantum fluctuations determine the spin-flop transition in hematite} 

\author{Tobias Dannegger} 
\email{Contact author: tobias.dannegger@uni-konstanz.de} 
\affiliation{\affilUKN} 

\author{Imre Hagymási} 
\affiliation{\affilWRC} 

\author{Levente Rózsa} 
\affiliation{\affilWRC} 
\affiliation{\affilBME} 

\author{Ulrich Nowak} 
\affiliation{\affilUKN} 

\date{\today} 

\begin{abstract} 
Magnetic phase transitions between ordered phases are often understood on the basis of semi-classical spin models. 
Deviations from the classical description due to the quantum nature of the atomic spins as well as quantum fluctuations are usually treated as negligible if long-range order is preserved, and are rarely quantified for actual materials. 
Here, we demonstrate that a fully quantum-mechanical framework is required for a quantitatively correct description of the spin-flop transition in the insulating altermagnet hematite between the collinear antiferromagnetic and the weakly ferromagnetic spin-flop phase at low temperature. 
By applying both exact diagonalization and density-matrix renormalization group theory to the quantum Heisenberg Hamiltonian, we show how a quantum-mechanical treatment of an ab initio parametrized spin model can significantly improve the predicted low-temperature spin-flop field over a classical description when compared to measurements. 
Our results imply that quantum fluctuations have a measurable influence on selecting the ground state of a system out of competing ordered magnetic phases at low temperature. 
\end{abstract} 

\maketitle

\ifusesections\section{Introduction}\fi 

Magnetic ordering is inherently a quantum-mechanical phenomenon, which relies on concepts like the spin angular momentum and the Pauli exclusion principle. 
However, ordered phases are characterized by a finite local expectation value of the spin operator, and this local magnetization vector may also be described by the methods of classical statistical mechanics or field theory. 
Common approaches for treating such systems include classical generalized Heisenberg models on the atomic level and continuum micromagnetic models on the mesoscopic scale. 
Density-functional theory also utilizes the local magnetization density, thus it is particularly well suited for parametrizing such classical models~\cite{Antropov1996,Rohrbach2004, Mazurenko2005, Logemann2017, Pozun2011, HematiteAbInitio}. 
The frequencies of magnetic excitations may be determined within the classical approach, enabling the comparison with (anti)ferromagnetic resonance~\cite{Artman1965, Morrison1973} or neutron scattering measurements~\cite{Samuelsen1970}, which provide an alternative way of fitting the parameters of the model. 
The numerical solution of these models using stochastic atomistic spin dynamics~\cite{Evans2014_Review} or micromagnetic simulations~\cite{Atxitia2017_FundamentalsLLB, Fidler2000} has proven immensely successful in uncovering and interpreting exotic magnetic textures and dynamical phenomena in the field of spintronics. 

However, the classical approach cannot reproduce certain phenomena connected to the quantum nature of the magnetic excitations, or magnons, in these ordered phases. 
The Bose--Einstein statistics of magnons results in the magnetization of ferromagnets scaling as $T^{\frac{3}{2}}$ at low temperature~\cite{Bloch1930}, in contrast to the linear temperature dependence predicted by the classical theory~\cite{WatsonBlumeVineyard}. 
In antiferromagnets, quantum fluctuations of the magnons lead to a reduction in the sublattice magnetizations compared to the saturation value~\cite{Anderson1952}, and these fluctuations also split the degeneracy via an order-by-disorder mechanism between states that classically have the same energy~\cite{Jolicoeur1990}. 
Recent efforts endeavored to correct these shortcomings of classical simulations by applying a quantum thermostat~\cite{Woo2015,BarkerBauer2019_QuantumThermostat} or via rescaling the temperature~\cite{Evans2015}. 

\begin{figure} 
\centering\includegraphics{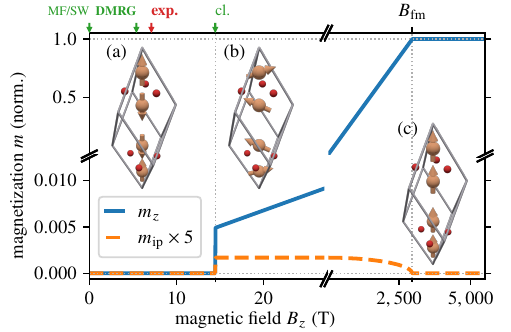} 
\caption{Out-of-plane ($m_z$) and in-plane ($m_{\mathrm{ip}}$) magnetization of hematite as a function of a magnetic field applied along the $z$-axis, calculated analytically at $T = 0$ within classical theory. 
The insets show the corresponding spin configurations within the primitive unit cell: (a) antiferromagnetic ground state, (b) spin-flop phase, (c) ferromagnetically polarized phase. 
The green arrows show the spin-flop field in the classical theory compared to mean-field (\acronym{MF}) and spin-wave theory (\acronym{SW}) calculations and \acronym{DMRG} results presented below. 
The red arrow shows the experimentally observed value. 
} 
\label{fig:Phases} 
\end{figure}

Recently it was suggested that quantum fluctuations are also responsible for the discrepancy in the spin-flop fields between experiments and classical atomistic spin simulations in the insulating iron oxide hematite (\ce{\alpha-Fe2O3})~\cite{HematiteAbInitio}. 
At low temperature, hematite assumes a collinear antiferromagnetic order with spins of all four sublattices aligned along the $z$-axis of threefold rotational symmetry, as depicted in \fig~\ref{fig:Phases}(a). 
If an external field is applied along the $z$-axis, the spin system undergoes a first-order transition into the spin-flop phase depicted in \fig~\ref{fig:Phases}(b)~\cite{Lebrun2019}. 
In addition to the magnetization along the field direction, the system acquires a small in-plane magnetization due to canting between the sublattices induced by the Dzyaloshinskii-Moriya interaction (\acronym{DMI})~\cite{Dzyaloshinskii1958, Moriya1960a, Moriya1960b}. 
For high enough (experimentally unachievable) fields, the system would become completely polarized along the field direction in a second-order transition, \fig~\ref{fig:Phases}(c). 
The ground state also transforms into the spin-flop or weak ferromagnetic state with \acronym{DMI}-induced canting when the temperature is raised above the Morin temperature $T_\text{M} \approx \qty{250}{\kelvin}$~\cite{Morin1950} in the field-free case. 

The classical atomistic spin dynamics simulations based on an \textit{ab initio}-parametrized spin model in Ref.~\cite{HematiteAbInitio} reproduced the experimentally observed Morin and Néel temperatures accurately, but overestimated the spin-flop field at zero temperature (cf.~\fig~\ref{fig:Phases}). 
This is caused by a qualitative difference in the low-temperature behavior: below \qty{160}{\kelvin}, the spin-flop field increases approximately linearly as the temperature is decreased in classical simulations, reaching \qty{14.5}{\tesla} at $T = 0$, while in experiments it stays approximately constant around \qty{7}{\tesla}~\cite{HematiteAbInitio, Lebrun2019}. 
It was demonstrated in Ref.~\cite{HematiteAbInitio} that a mean-field description of a quantum instead of a classical spin model can qualitatively reproduce the weak temperature dependence of the spin-flop field observed in the experiments. 
However, the mean-field model proved to be quantitatively inaccurate in its prediction of the critical temperatures or fields. 
Determining the spin-flop field of quantum magnets quantitatively is challenging from the theoretical point of view: in two-sublattice antiferromagnets without \acronym{DMI}, predictions exist based on spin-wave theory at zero temperature~\cite{Wang1964}, but this has proven difficult to extend to finite temperatures~\cite{Anderson1964}. 
In the presence of \acronym{DMI}, calculations have been mainly restricted to mean-field theory~\cite{Berger1973}. 

Here, we present a quantum spin model for hematite based on the material-specific description of Ref.~\cite{HematiteAbInitio}. 
We transform the parameters of the classical model to the quantum one while keeping the spin-wave frequencies constant. 
Within linear spin-wave theory, this model does not reproduce the correct ground state, but we will show that a more accurate treatment of quantum fluctuations in exact diagonalization (\acronym{ED}) and density-matrix renormalization group (\acronym{DMRG}) theory restores the collinear antiferromagnetic ground state, and predicts a spin-flop field at zero temperature that compares favorably with the experimental value (see \fig~\ref{fig:Phases}).

\ifusesections\section{The quantum-mechanical spin model} \label{sec:QuantumSpinModel}\fi 

The magnetic properties of hematite can be understood on the basis of an extended Heisenberg model with the Hamiltonian~\cite{HematiteAbInitio} 
\begin{equation} 
  \hamil = -\tfrac{1}{2} \sum_{i\neq j} \vec{S}_i^T \tensorial{J}_{ij} \vec{S}_j 
    - \sum_i \bigl[ d_2 S_{i,z}^2 + d_4 S_{i,z}^4 + \mu \vec{B} \cdot \vec{S}_i \bigr] 
  \label{eq:ClassicalHamiltonian} 
  . 
\end{equation} 
The $\tensorial{J}_{ij}$ represent tensorial exchange interactions, incorporating the isotropic exchange, \acronym{DMI}, and easy-plane two-ion anisotropies. 
The latter lead to the Morin transition in competition with the easy-axis on-site anisotropy energies $d_2$ and $d_4$. 
The fourth-order term creates an additional energy barrier between the collinear and the canted states and needs to be included to correctly represent the nature of the Morin and spin-flop transitions as first-order phase transitions. 
The last term is the Zeeman energy with the atomic magnetic moment~$\mu$ and the external magnetic field~$\vec{B}$. 

Classically, the $\vec{S}_i$ are seen as vectors on the unit sphere. 
For the quantum-mechanical treatment, we have to replace them with angular momentum operators $\vecop{S}_i = (\op{S}_{i,x}, \op{S}_{i,y}, \op{S}_{i,z})^T$. 
To express the Hamiltonian in matrix form, we replace the $x$- and $y$-components of the spin operators with the ladder operators $\op{S}_i^{\pm} = \op{S}_i^x \pm \iu \op{S}_i^y$ (cf.~\cite{SM}). 
The classical model is obtained in the $S\rightarrow\infty$ limit of the quantum description, where the components of the vector operator $\vecop{S}_{i}/S$ assume infinitely many eigenvalues between $-1$ and $1$, and the three components together map the whole unit sphere. 
To establish a correspondence between the classical and quantum Hamiltonians, we rescale terms that are of second order in spin operators by a factor of $S^2$ and fourth-order terms by $S^4$: 
\begin{equation} 
  \tensorial{J}_{ij}^{\textrm{q}}(S) = \frac{\tensorial{J}^{\textrm{c}}_{ij}}{S^2},\; 
  d^{\textrm{q}}_2(S) = \frac{d^{\textrm{c}}_2}{S^2},\; 
  d^{\textrm{q}}_4(S) = \frac{d^{\textrm{c}}_4}{S^4}; 
  \label{eq:ScalingSquare} 
\end{equation} 
where the q and c superscripts denote the coefficients in the quantum and the classical models, respectively. 
With this approach, the spin-wave frequencies calculated from performing a linearized Holstein--Primakoff transformation on the quantum Hamiltonian and from the linearization of the classical Landau--Lifshitz--Gilbert equation will be identical. 
This rescaling was used in Ref.~\cite{BarkerBauer2019_QuantumThermostat}, where the spin-wave spectrum was compared to previous experimental results. 

The atomic magnetic moments in hematite appear due to the 3d$^5$ electron configuration of the \ce{Fe^3+} ions. 
As free ions, these should have a quantum number of $S = \tfrac{5}{2}$ 
according to Hund's rules, corresponding to a magnetic moment of approximately $5 \mu_\text{B}$. 
For the oxygen-ligated ions in hematite, however, \acronym{DFT} calculations have shown that delocalization of the 3d electrons leads to a reduction of the atomic moment toward $4 \mu_\text{B}$ ($4.11$ to $4.24 \mu_\text{B}$)~\cite{Rohrbach2004, Pozun2011, HematiteAbInitio}, which is confirmed by experimental measurements~\cite{Hill2008}. 
This would suggest an alternative quantum number of $S = 2$. 
In the following, we will therefore present results with both of those values, $S = 2$ and $S = \tfrac{5}{2}$.

\ifusesections\section{Exact diagonalization} \label{sec:ED}\fi 

\begin{figure} 
\centering\includegraphics{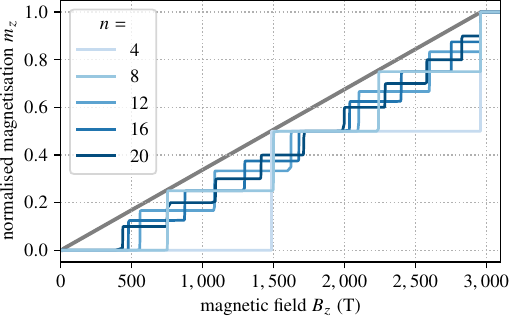} 
\caption{Magnetization curve for a system of $n$ spins with a quantum number $S = \tfrac{1}{2}$ computed with \acronym{ED}. 
The thicker gray line indicates the classical magnetization curve. 
} 
\label{fig:EdMagCurve} 
\end{figure}

To determine the ground state of the quantum spin Hamiltonian exactly, we express it in matrix form using the basis of products of $S_{i}^{z}$ eigenstates with projection quantum numbers $m_i = -S, \ldots, S$ for each spin $i \in \{1, \ldots, n\}$. 
For $n$ spins, this basis consists of $N = (2 S + 1)^n$ states. 
Since the primitive magnetic unit cell contains four spins, the total number~$n$ of spins in the system should be a multiple of four. 
To have the same number of neighbors regardless of the system size, we apply periodic boundary conditions. 
To find the ground state, i.e.,\ the lowest-energy eigenstate of $\op{\hamil}$, we use the Arnoldi method (as implemented in the \acronym{ARPACK} package~\cite{ArpackManual}), which is a matrix-free iterative method that can compute the lowest eigenvalue and corresponding eigenvector with a runtime complexity of $\mathcal{O}(N^2)$. 
To investigate the spin-flop transition, we apply an external magnetic field $\vec{B} = (0, 0, B_z)$, determine the resulting ground state vector $\vec{\psi}_0(B_z)$ and calculate the expectation value of the normalized magnetization in $z$-direction, which is given by 
\begin{equation} 
  m_z(B_z) = \frac{1}{\hbar Sn} \sum_{i=1}^n \vec{\psi}_0^\dagger \op{S}_{i}^{z} \vec{\psi}_0 
  . 
\end{equation} 
Since at the spin-flop field $m_{z}$ displays a step in the thermodynamic limit (cf.\ \fig~\ref{fig:Phases}), we identify the spin-flop field in a finite-sized system with the position of the first such step in $m_{z}$. 
To keep the atomic magnetic moment consistent with the classical case, we set $\mu_{\text{q}} S=\mu_{\text{cl}}$ in the Zeeman term for each $S$ value,  where $\mu_{\text{cl}} = \num{4.2313} \mu_{\text{B}}$ is the magnetic moment per iron atom from the \textit{ab initio} calculations~\cite{HematiteAbInitio}. 
An example of the resulting magnetization curves, here for the quantum number $S = \tfrac{1}{2}$, is shown in \fig~\ref{fig:EdMagCurve} for magnetic fields from zero to above \qty{3000}{\tesla}, at which point the system becomes fully polarized. 
Below the transition into this ferromagnetic-like state, we see that the magnetization of the quantum-mechanical system consistently remains below the classical magnetization curve, suggesting that quantum fluctuations reduce the polarization in the spin-flop phase. 
This is consistent with the observation that the classical system reaches its saturation magnetization at a slightly smaller critical field than the quantum model (the difference is less than \qty{2}{\tesla} and therefore not visible in the figure). 

Since the terms in the Hamiltonian that do not conserve the total spin $S^{z} = \frac{1}{N} \sum_{i} S_{i}^{z}$ are small, $m_{z}$ is almost a conserved quantity and it increases in steps. 
The position of the first step, identified with the spin-flop field, is very large in very small systems and slowly converges toward the bulk value for increasing system sizes, as illustrated in \fig~\ref{fig:EdMagCurve}. 
Following Ref.~\cite{Xu2019}, we expect this finite-size correction to be proportional to the inverse volume $n^{-1}$, which we use to extrapolate to the corresponding bulk value. 

\begin{figure} 
\centering\includegraphics{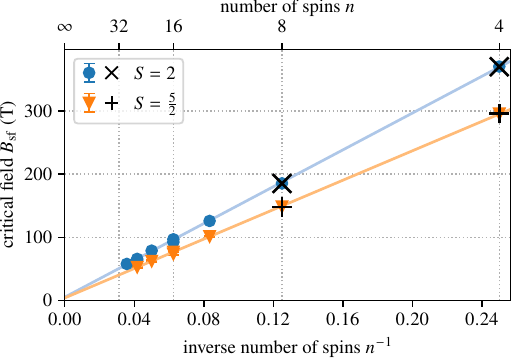} 
\caption{Spin-flop field computed by \acronym{ED} and \acronym{DMRG} for clusters of different sizes~$n$ and quantum numbers~$S$. 
The colored points are \acronym{DMRG} results and the black crosses mark the corresponding \acronym{ED} results for the two smallest clusters. 
Uncertainty ranges for the \acronym{DMRG} results were also calculated but are smaller than the symbol size. 
The two lines are linear fits to the chain-like clusters used to extrapolate the critical field in the bulk limit. 
} 
\label{fig:Dmrgspin-flop} 
\end{figure}

Since the computational effort of \acronym{ED} scales with $(2 S + 1)^{2n}$, these calculations are limited to very small clusters, even with further optimizations. 
Due to the long-range interactions in hematite, the memory requirements are also significant, limiting our \acronym{ED} calculations to systems of 8 spins for the relevant quantum numbers of $S = 2$ and $\tfrac{5}{2}$. 
The spin-flop fields obtained for these small systems are shown as black crosses in \fig~\ref{fig:Dmrgspin-flop}. 
While two data points would suffice for a linear extrapolation, it does not allow us to quantify the statistical uncertainty of the extrapolated values. 
We therefore turn to \acronym{DMRG} theory to scale our calculations to larger clusters and obtain more data points. 

\ifusesections\section{Density-matrix renormalization group calculations} \label{sec:DMRG}\fi 

The \acronym{DMRG} algorithm~\cite{White1992,White1993,Schollwöck2005_Review,Schollwöck2011_Review,Hallberg2006_Review} is a variational technique, which relies on the assumption that the many-body wave function can be represented faithfully in a matrix-product-state form. 
The accuracy of the ansatz is controlled by the bond dimension (size of the matrices), which is gradually increased during the local \acronym{DMRG} updates. 
We used the single-site variant of the \acronym{DMRG} algorithm with subspace expansion~\cite{Hubig2015}. 
The large sizes of the local Hilbert space for $S=2$ and $S = \tfrac{5}{2}$ and the lack of spin symmetries of the Hamiltonian severely impact the achievable system sizes in \acronym{DMRG}. 
Accordingly, the largest bond dimension is restricted to $\chi = \num{2500}$, since we have to work with full dense complex matrices. 
Nevertheless, this was sufficient to obtain converged results for clusters containing up to seven unit cells (28~spins). 
We then determined the spin-flop transition in the same fashion as for the \acronym{ED}. 

The \acronym{DMRG} results are also shown in \fig~\ref{fig:Dmrgspin-flop}. 
Reassuringly, the \acronym{ED} and \acronym{DMRG} results agree almost perfectly. 
For some system sizes, differently shaped clusters with the same number of spins are possible to simulate using \acronym{DMRG}, for example with six unit cells we considered both the shape $1 \times 1 \times 6$ and $1 \times 2 \times 3$. 
As can be seen in \fig~\ref{fig:Dmrgspin-flop}, the differences in the resulting spin-flop field between differently shaped clusters are relatively small (\qty{5}{\percent} for $n = 16$ and \qty{3}{\percent} for $n = 24$). 
For the extrapolation to the bulk case, we limit ourselves to the linear (chain-like) clusters of shape $1 \times 1 \times n$. 
Linear extrapolation with these clusters results in the bulk spin-flop fields $B_{\text{sf}}(S = 2) = \qty{5.4(5)}{\tesla}$ and $B_{\text{sf}}(S = \tfrac{5}{2}) = \qty{4.2(2)}{\tesla}$. 
This is a significant reduction compared to the value of \qty{14.5}{\tesla} predicted by the classical model and even slightly lower than the low-temperature experimental value of approximately \qty{7}{\tesla}~\cite{HematiteAbInitio}, suggesting that a quantum-mechanical treatment of the same spin model can indeed resolve the quantitative overestimation of the zero-temperature spin-flop field produced by semi-classical simulations.

\ifusesections\section{Discussion and comparison} \label{sec:Discussion}\fi

To estimate the importance of the quantum fluctuations captured by DMRG, we compare the obtained spin-flop fields to semi-analytical methods which are closer to the classical description, namely mean-field theory and linear spin-wave theory (further details in~\cite{SM}). 

\begin{figure} 
\centering\includegraphics{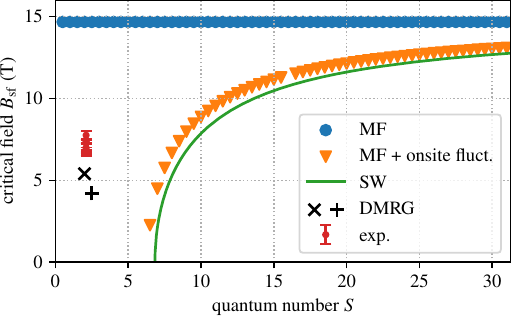} 
\caption{Spin-flop field for different quantum numbers calculated from mean-field theory (\acronym{MF}), both without and including on-site fluctuations, as well as spin-wave theory (\acronym{SW}), compared to the values extrapolated from the \acronym{DMRG} calculations and to experimental observations~\cite{HematiteAbInitio}. 
} 
\label{fig:CorrelatedCorrection} 
\end{figure}

In a completely mean-field approach, the way we scaled the parameters leads to a constant spin-flop field independent of the quantum number (blue points in \fig~\ref{fig:CorrelatedCorrection}), coinciding with the classical case. 
However, a correct description of the Morin transition requires a slightly modified approach~\cite{HematiteAbInitio}, where the on-site anisotropy terms are not included in the mean field. 
This allows for quantum fluctuations in the spin-flop phase, which lowers its energy and therefore reduces the spin-flop field significantly (orange points in \fig~\ref{fig:CorrelatedCorrection}). 
For $S \le 6$, the spin-flop field vanishes, and the flop state becomes the field-free ground state. 

Linear spin-wave theory is based on an expansion around the classical ground state of the system, but by performing the Bogoliubov transformation it also includes zero-point quantum fluctuations of magnons which reduce the energy of the system. 
These fluctuations are correlated between the sites, unlike the mean-field approach extended by on-site fluctuations discussed above; therefore, spin-wave theory is expected to approximate the quantum-mechanical ground state better. 
Performing Bogoliubov transformation on the magnon operators leads to a better approximation of the quantum-mechanical ground state that includes zero-point quantum fluctuations which reduce the energy of the system. Considering the correlations described by the zero-point fluctuations results in a slight decrease of the spin-flop field (green line in \fig~\ref{fig:CorrelatedCorrection}), compared to the mean-field result including on-site fluctuations. 
Comparing the results for different quantum numbers, we find that zero-point quantum fluctuations stabilize the spin-flop phase over the antiferromagnetic state in linear spin-wave theory. 
This may be connected to the fact that the magnon spectrum in the spin-flop phase contains a Goldstone mode due to the rotational freedom of the state around the field direction, while the spectrum in the collinear antiferromagnetic state is gapped by the anisotropy terms (the very small triaxial in-plane anisotropy is neglected in the model). 
The magnon states closer to zero energy in the spin-flop phase contribute strongly to the ground-state quantum fluctuations, thereby decreasing the energy of this state relative to the collinear phase. 
As in mean-field theory, the spin-flop field vanishes for small spin quantum numbers. 
This underscores the need for more exact methods such as \acronym{ED} and \acronym{DMRG} which can more accurately treat spin fluctuations, resulting in a reduced but finite spin-flop field for $S = 2$ and~$\tfrac{5}{2}$.

\ifusesections\section{Conclusions} \label{sec:Conclusions}\fi 

\ifusesections We\else To summarize, we\fi{} have shown that the highly accurate description of quantum fluctuations provided by \acronym{ED} and \acronym{DMRG} calculations is required to quantitatively determine the spin-flop field of hematite at low temperature. 
A classical spin model with \textit{ab initio} parameters can correctly describe the spin-flop field and the Morin transition above $T = \qty{160}{\kelvin}$, but it overestimates the spin-flop field at low temperatures compared to the measured value~\cite{HematiteAbInitio}. 
Transforming the parameters to a quantum model, we found that the spin-flop field vanishes at low-temperature in mean-field theory and linear spin-wave theory, which only limitedly account for quantum fluctuations. 
In contrast, the \acronym{DMRG} calculations result in a finite spin-flop field by correctly describing the fluctuations which stabilize the collinear antiferromagnetic phase. 
Extrapolating the calculated spin-flop field to the thermodynamic limit results in a value which compares favorably to the measurements~\cite{HematiteAbInitio}. 
The validity of the \acronym{DMRG} results is supported by the fact that they are in perfect agreement with \acronym{ED} at the system sizes that are achievable with the latter. 

Our results demonstrate that quantum fluctuations play an important role for the quantitatively correct description of phase transitions, not only in hematite, but most likely in other insulating magnets as well. 
This is observed in spite of the transition happening between two long-range-ordered phases and the not particularly small ($S = 2$ or $S = \tfrac{5}{2}$) spin magnetic moment of the ferrous ions in hematite, both of which are conventional arguments for supporting a semi-classical treatment of the spin model. 
This implies that it may be necessary to go beyond the semi-classical description when parametrizing spin models based on first-principles calculations for the quantitative interpretation of low-temperature measurements, particularly in insulating magnets with even smaller quantum numbers.

\begin{acknowledgments} 
We are grateful to H. Lange for her suggestions and helpful discussions about \acronym{ED} and \acronym{DMRG}. 
This work has been supported by the German Research Foundation (DFG) under project No.~423441604. 
TD and UN acknowledge additional support from the DFG through project No.~425217212 (SFB~1432, project B06). 
The \acronym{ED} calculations were supported by local computing resources through the core facility \acronym{SCCKN}. 
IH was supported by the Hungarian National Research, Development and Innovation Office (NKFIH) through Grants No.~FK142985 and No.~K134983, by the Quantum Information National Laboratory of Hungary, as well as by the János Bolyai Research Scholarship of the Hungarian Academy of Sciences. 
IH acknowledges the use of the Noctua2 cluster at the Paderborn Center for Parallel Computing (PC$^2$). 
LR gratefully acknowledges support by the National Research, Development, and Innovation Office (NRDI) of Hungary under Project No. FK142601, by the Ministry of Culture and Innovation and the National Research, Development and Innovation Office within the Quantum Information National Laboratory of Hungary (Grant No.\ 2022-2.1.1-NL-2022-00004), and by the Hungarian Academy of Sciences via a János Bolyai Research Grant (Grant No.\ BO/00178/23/11). 
\end{acknowledgments}

\end{document}